\documentclass[aps,prd,twocolumn,floatfix,nofootinbib,preprintnumbers,superscriptaddress]{revtex4-1}

\usepackage[utf8]{inputenc}
\usepackage{newtxtext}
\usepackage[libertine]{newtxmath}

\usepackage{graphicx}
\usepackage{slashed}
\usepackage{amsfonts}
\usepackage{changes}

\usepackage[english]{babel}

\usepackage{blindtext}
\usepackage{microtype}

\usepackage[colorlinks,linkcolor=blue,anchorcolor=blue,citecolor=blue,urlcolor=blue,breaklinks=true]{hyperref}

\def\be{\begin{equation}}
\def\ee{\end{equation}}
\def\bea{\begin{eqnarray}}
\def\eea{\end{eqnarray}}

\begin{document}

\preprint{ADP-21-20/T1167}

\title{Constraints on the dark photon from deep inelastic scattering}

\author{A. W.~Thomas}
\affiliation{ARC Centre of Excellence for Dark Matter Particle Physics and CSSM, Department of Physics, University of Adelaide, SA 5005, Australia}
\date{\today}

\author{X. G.~Wang}
\affiliation{ARC Centre of Excellence for Dark Matter Particle Physics and CSSM, Department of Physics, University of Adelaide, SA 5005, Australia}

\author{A. G.~Williams}
\affiliation{ARC Centre of Excellence for Dark Matter Particle Physics and CSSM, Department of Physics, University of Adelaide, SA 5005, Australia}

\begin{abstract}
We investigate constraints on the dark photon arising from an analysis of deep inelastic scattering (DIS) data.
We perform extractions of parton distribution functions (PDFs) with and without a dark photon being present and allow the dark photon mixing parameter and mass to vary. We also include the effects of vector meson dominance to ensure the correct photo-production limit. By considering the variation of the total $\chi^2$ arising from such extractions we infer exclusion limits on the kinetic mixing parameter of the dark photon for dark photon mass up to $80\ {\rm GeV}$.
\end{abstract}


\maketitle

\section{Introduction}
Dark matter (DM) is thought to make up $85\%$ mass of the Universe
and its existence has been confirmed in a multitude of ways~\cite{Planck:2015fie, Abazajian:2014fta}. 
Understanding the nature and origin of DM is one of the most important challenges in particle physics and astronomy today.
While the direct detection of weakly interacting massive particles (WIMPs) is an area of very active investigation~\cite{SABRE:2018zhm, Amare:2021yyu, COSINE-100:2021xqn}, 
the stringent limits from null experiments~\cite{XENON:2017lvq, Bondarenko:2019vrb, DEAP:2020iwi} have motivated a number of alternative hypotheses for the nature of DM.
In recent years, the dark photon as a portal of interactions between DM and Standard Model (SM) particles has been receiving considerable attention. For a review of the current status of theoretical and experimental studies on the dark photon see Ref.~\cite{Fabbrichesi:2020wbt}. 

The dark photon was firstly proposed as an extra $U(1)$ gauge boson~\cite{Fayet:1980ad, Fayet:1980rr} that can be either massless or it may acquire a mass.
Currently, the massive dark photon is of particular interest as it is more
readily accessible in experimental searches. 
It interacts with SM particles through kinetic mixing with hypercharge~\cite{Okun:1982xi}
\begin{equation}
\mathcal{L}  \supset - \frac{1}{4} F'_{\mu\nu} F'^{\mu\nu} + \frac{\bar{m}^2_{A'}}{2} A'_{\mu} A'^{\mu} + \frac{\epsilon}{2 \cos\theta_W} F'_{\mu\nu} B^{\mu\nu} \, .
\end{equation}
Here $\theta_W$ is the Weinberg angle, $F'_{\mu\nu}$ is the dark photon strength tensor and $\epsilon$ is the mixing parameter quantify the mixing of the dark photon with the $B$-boson of the standard electroweak theory.

There have been numerous experimental searches for the dark photon~\cite{Merkel:2014avp, LHCb:2019vmc, CMS:2019buh, BaBar:2017tiz, Banerjee:2019pds}.
The strongest constraints come from the NA64~\cite{Banerjee:2019pds, Andreev:2021fzd} experiment for $1\ {\rm MeV} \le \bar{m}_{A'} \le 250\ {\rm MeV}$ 
and from the BaBar~\cite{BaBar:2017tiz} experiment for $250\ {\rm MeV} \le \bar{m}_{A'} \le 8\ {\rm GeV}$, 
in which the missing-energy events are described as invisible decays of the dark photon, leading to $\epsilon \le {\cal O}(10^{-3})$.
These constraints could be weakened by taking into account the detailed structure of the dark sector~\cite{Essig:2009nc}. 

So-called ``decay-agnostic" processes can also provide constraints on the dark photon. These are independent of the details of its production mechanism and decay modes.
Such bounds can be obtained from measurements of the electroweak precision observables (EWPO)~\cite{Hook:2010tw, Curtin:2014cca} and the muon $g-2$~\cite{Pospelov:2008zw}.
Another ``decay-agnostic" process is $e^{\pm} p$ deep inelastic scattering (DIS), 
which can potentially probe dark photons with masses up to 100 TeV because of the broad energy reach of experimental measurements~\cite{LHeCStudyGroup:2012zhm}. 
Recently, a competitive constraint on the dark photon mixing parameter was derived from DIS using data from HERA~\cite{Kribs:2020vyk}.
The dark photon contribution to the proton structure function leads to non-Dokshitzer-Gribov-Lipatov-Altarelli-Parisi (non-DGLAP) scaling violations, which were proposed as a smoking gun relevant for future experiments~\cite{Kribs:2020vyk}.
 
However, in Ref.~\cite{Kribs:2020vyk} the parton distribution functions (PDFs) used were from the best fit results of HERA analysis without consideration of the possibility of a dark photon. When examining the consequences of the addition of a dark photon it is desirable to extract the PDFs from the data taking into account the changes in the PDFs which result from including the dark photon contribution to the scattering.
The necessity of a simultaneous determination on the PDFs and beyond SM effects was also demonstrated in Ref.~\cite{Carrazza:2019sec}, by considering representative four-fermion operators in the SM effective field theory.
Moreover, in analyzing DIS data, one should also take into account the hadronic fluctuations of the SM photon~\cite{Levy:1997fe}, as well as higher twist effects~\cite{Accardi:2016qay},
both of which also have non-DGLAP features.  

In this work, we investigate the properties of the dark photon that are consistent with data from the DIS process, while allowing variations in the PDFs required in the presence of the dark photon. 
Our analysis then places upper limits on the allowed kinetic mixing of the dark photon as a function of the dark photon mass.
We employ a two-component model of DIS, taking into account the transition of the SM photon to a non-perturbative $q\bar{q}$ pair,  using the vector meson dominance (VMD) framework. The VMD contributions need to be included as otherwise we would be missing important contributions that affect the parameter space of the dark photon for masses below 10~GeV.
By refitting the DIS data in this way, we place upper limits on the mixing parameter of the dark photon for masses up to $80\ {\rm GeV}$.

\section{Proton structure function}
\label{}
With the inclusion of the dark photon contribution, the transverse structure function of the proton is given by~\cite{Kribs:2020vyk}
\begin{equation}
\label{eq:F2tilde}
\tilde{F}_2 = \sum_{i,j=\gamma,Z,A_D} \kappa_i \kappa_j F_2^{ij}
\, ,
\end{equation}
where $\kappa_i = Q^2/(Q^2 + M_{V_i}^2)$. At leading order (LO) in $\alpha_s$ 
\begin{equation}
F_2^{ij} = \sum_q x f_q (C^v_{i,e} C^v_{j,e} + C^a_{i,e} C^a_{j,e}) (C^v_{i,q} C^v_{j,q} + C^a_{i,q} C^a_{j,q}) \, , 
\end{equation}
where $x$ is the Bjorken variable and $f_q$ are the PDFs of the quark flavors $q = u, \bar{u}, d, \bar{d}, c, \bar{c}, s, \bar{s}, b, \bar{b}$.

The couplings to the electron and quarks for the photon are
\begin{equation}
\{ C_{\gamma, e}^v, C_{\gamma, u}^v, C_{\gamma, d}^v \} = \left\{ -1, \frac{2}{3}, - \frac{1}{3} \right\}, \ \ \ C_{\gamma}^a = 0 
\, ,
\end{equation}
while for the unmixed Z boson they are
\begin{equation}
\bar{C}_Z \sin 2\theta_W = T_3^f - 2 q_f \sin^2\theta_W, \ \ \ \bar{C}_Z^a \sin 2\theta_W = T_3^f \, ,
\end{equation}
where 
\begin{equation}
\{T_3^e, T_3^u, T_3^d \} = \left\{ -\frac{1}{2}, \frac{1}{2}, -\frac{1}{2} \right\},\ \{ q_e, q_u, q_d\} = \left\{ -1, \frac{2}{3}, -\frac{1}{3} \right\} \, .
\end{equation}

After diagonalizing the mixing term through field redefinitions, the couplings of the physical $Z$ and $A_D$ to SM particles are
given by \cite{Kribs:2020vyk}
\begin{eqnarray}
C_Z^v &=& (\cos\alpha - \epsilon_W \sin\alpha) \bar{C}_Z^v + \epsilon_W \sin\alpha \cot \theta_W C_{\gamma}^v ,\nonumber\\
C_Z^a &=& (\cos\alpha - \epsilon_W \sin\alpha) \bar{C}_Z^a 
\end{eqnarray}
and 
\begin{eqnarray}
C_{A_D}^v &=& - (\sin\alpha + \epsilon_W \cos\alpha) \bar{C}_Z^v + \epsilon_W \cos\alpha \cot \theta_W C_{\gamma}^v ,\nonumber\\
C_{A_D}^a &=& - (\sin\alpha + \epsilon_W \cos\alpha) \bar{C}_Z^a 
\, .
\end{eqnarray}
Here $\alpha$ is the $\bar{Z}-A'$ mixing angle
\begin{eqnarray}
\tan \alpha &=& \frac{1}{2\epsilon_W} \Big[ 1 - \epsilon^2_W - \rho^2 \nonumber\\
&& - {\rm sign}(1-\rho^2) \sqrt{4\epsilon_W^2 + (1 - \epsilon_W^2 - \rho^2)^2} \Big] \, , 
\end{eqnarray}
with
\begin{eqnarray}
\epsilon_W &=& \frac{\epsilon \tan \theta_W}{\sqrt{1 - \epsilon^2/\cos^2\theta_W}} ,\nonumber\\
\rho &=& \frac{\bar{m}_{A'}/\bar{m}_{\bar{Z}}}{\sqrt{1 - \epsilon^2/\cos^2\theta_W}} \, .
\end{eqnarray}
The physical masses of the Z boson and the dark photon are
\begin{eqnarray}
m^2_{Z, A_D} &=& \frac{m_{\bar{Z}}^2}{2} [ 1 + \epsilon_W^2 + \rho^2 \nonumber\\
&& \pm {\rm sign}(1-\rho^2) \sqrt{(1 + \epsilon_W^2 + \rho^2)^2 - 4 \rho^2} ] \, .
\end{eqnarray}

Given that the dark photon contribution is expected to be small, an accurate determination of its parameters relies on refined analysis of the DIS data.
At low $Q^2$, and hence low resolution, the interaction of the virtual photon with the nucleon is not well described in the parton model. 
For example, it does not ensure the constraint of current conservation in the photo-production limit, namely that the proton structure function be proportional to $Q^2$ as $Q \rightarrow 0$. Instead, following Badelek and Kwiecinski~\cite{Kwiecinski:1987tb, Badelek:1992ua} this is naturally ensured by using a two-component model of nucleon structure functions which  incorporates vector meson dominance (VMD) at low-$Q^2$. In that model the photon acts on the whole target like a virtual vector meson. Such a model was applied to phenomenological analyses of DIS in Refs.~\cite{Martin:1998dr, Szczurek:1999rd, Wang:2019tmb}.
Embedding Eq.~(\ref{eq:F2tilde}) into this model, the proton structure function can be written as
\begin{equation}
\label{eq:F2-BK}
F_2(x,Q^2) = F_2^{\rm VMD}(x,Q^2) + \frac{Q^2}{Q^2 + M_0^2} \tilde{F}_2(\bar{x},Q^2+M_0^2) \, ,
\end{equation}
where
\begin{equation}
\bar{x} = x \frac{Q^2 + M_0^2}{Q^2 + x M_0^2} \, .
\end{equation}
The parameter $M_0$, which lies in the range 
$1.0-1.5\ {\rm GeV}^2$~\cite{Martin:1998dr}, controls a smooth transition from the VMD to the partonic regime.
The VMD term has the form
\begin{equation}
F_2^{\rm VMD} = \frac{Q^2}{\pi} \sum_V \frac{M_V^4 \sigma_{VN}}{f_V^2 (Q^2 + M_V^2)^2} \Omega(x,Q^2) \, ,
\end{equation}
where $V = \rho^0, \omega$ and $\phi$, and the photon-vector-meson coupling constants are 
\begin{equation}
\frac{f^2_V}{4\pi} = \frac{\alpha^2 M_V}{3 \Gamma_{V\rightarrow e^+ e^-}} \, ,
\end{equation}
equal to 2.28, 26.14, and 14.91 for $\rho^0$, $\omega$ and $\phi$, respectively. 
The vector meson-proton cross sections are
\begin{eqnarray}
\sigma_{\rho p} &=& \sigma_{\omega p} = \frac{1}{2} \Big[ \sigma(\pi^+ p) + \sigma(\pi^- p) \Big]\ ,\nonumber\\
\sigma_{\phi p} &=& \sigma( K^+ p) + \sigma(K^- p) - \frac{1}{2} \Big[ \sigma(\pi^+ p) + \sigma(\pi^- p) \Big] \, ,
\end{eqnarray}
where the cross sections are parametrized in the 
form~\cite{Donnachie:1992ny}
\begin{eqnarray}\label{eq:sigma-vp}
\sigma_{\rho p} &=& \sigma_{\omega p} = 13.63 s^{\epsilon} + 31.79 s^{-\eta}\ ,\nonumber\\
\sigma_{\phi p} &=& 10.01 s^{\epsilon} + 2.72 s^{-\eta} \, . 
\end{eqnarray}
Here $\epsilon=0.08$ and $\eta=0.45$ are taken from Regge theory  and the resulting cross sections are given in units of mb.

In phenomenological analysis, a Gaussian form factor is often  introduced~\cite{Szczurek:1999rd} 
\begin{equation}
\Omega(x,Q^2) = \exp(-(\Delta E/\lambda_G)^2) \, ,
\end{equation}
where $1/\Delta E$ characterizes the lifetime of the hadronic fluctuations of the photon. 
In the target reference frame,
\begin{equation}
\Delta E = \frac{M_V^2 + Q^2}{Q^2} \, M_N x \, .
\end{equation}

In this exploratory study we adopt the leading order (LO) HERA parametrization of the PDFs at an initial scale, $Q_0^2 = 1.9\ {\rm GeV}^2$~\cite{H1:2015ubc},
 \begin{eqnarray}
 x g (x,Q_0^2)   &=& A_{g} x^{B_g} (1-x)^{C_g} , \nonumber\\
 x u_v(x,Q_0^2) &=& A_{u_v} x^{B_{u_v}} (1-x)^{C_{u_v}} [1 + E_{u_v} x^2] , \nonumber\\
 x d_v(x,Q_0^2) &=& A_{d_v} x^{B_{d_v}} (1-x)^{C_{d_v}}  ,\nonumber\\
 x \bar{u}(x,Q_0^2) &=& A_{\bar{u}} x^{B_{\bar{u}}} (1-x)^{C_{\bar{u}}} [1 + D_{\bar{u}} x] ,\nonumber\\
 x \bar{d}(x,Q_0^2) &=& (1-0.4) A_{\bar{d}} x^{B_{\bar{d}}} (1-x)^{C_{\bar{d}}} ,\nonumber\\
  x \bar{s}(x,Q_0^2) &=& 0.4 A_{\bar{d}} x^{B_{\bar{d}}} (1-x)^{C_{\bar{d}}} \, .
 \end{eqnarray}
We evolve the PDFs to the DIS scale $Q^2$ using the APFEL program~\cite{Bertone:2013vaa}.
At LO, we use $\alpha_s(M_Z) = 0.13$. The heavy flavour thresholds are $m_c = 1.47\ {\rm GeV}$, $m_b = 4.5\ {\rm GeV}$, and $m_t = 173\ {\rm GeV}$.

\section{Results}
\label{sec:results}
In our analysis, we take the HERA data~\cite{H1:2015ubc} for the reduced cross section with $\sqrt{s} = 318\ {\rm GeV}$ and $Q^2 \in [3.5,30000]\ {\rm GeV}^2$.
We only include data points in the region $y \le 0.1$, so that the effect of the longitudinal structure function can be neglected.
We also include BCDMS~\cite{BCDMS:1989qop} data to constrain the structure function in the large $x$ region. 

Bearing in mind that we use a limited data set, here we find it necessary to constrain some parameters. 
Since the structure function is less sensitive to the $d$-quark distributions, we fix the $d_v$ and $\bar{d}$ distributions from the HERA analysis.
The parameters $A_{u_v}$ and $A_g$ are constrained by number and momentum sum rules, respectively, leaving 7 free parameters in the $x u_v$, $x \bar{u}$, and $xg$ distributions. 

We first fit the HERA and BCDMS data without the dark photon.
The fit results using Eqs.~(\ref{eq:F2tilde}) and (\ref{eq:F2-BK}) are given in Tab.~\ref{tab:refit-7para}, respectively. It is clear that the inclusion of VMD in the two-component model significantly improves the $\chi^2$.
The best fit corresponds to $\chi^2_{\rm d.o.f} = 291.79/(259- 8) = 1.16$ and $\lambda_G = 0.897\ {\rm GeV}$, with 
$M_0^2$ fixed at $1.0\ {\rm GeV}^2$.
We also show the second moments of $xg$ and $xu^+$ at $Q^2 = 4\ {\rm GeV}^2$, 
which are in agreement with the latest lattice determinations~\cite{Alexandrou:2020sml} and the unweighted average of the global fit analyses~\cite{Constantinou:2020hdm}.
\begin{table*}[!htbp]
 \begin{center}
\begin{tabular}{c|ccc|ccc}
\hline\hline
       \     &                                  \multicolumn{3}{c|}{without VMD}                        &     \multicolumn{3}{c}{with VMD ($\lambda_G = 0.897\ {\rm GeV}$)}      \\ \hline
       \     &              $xg$             &          $x u_v$             &            $x \bar{u}$      &                $xg$              &          $x u_v$                &           $x \bar{u}$          \\ \hline
        A   &             5.3368           &            4.4790            & $0.0894\pm 0.0049$  &              4.9008             &           4.4531                & $0.0904 \pm 0.0066$    \\ 
       B    & $0.0745\pm 0.0319$ & $0.7441\pm 0.0126$ & $-0.3020\pm 0.0099$ & $0.0659\pm 0.0442$   & $0.7419 \pm 0.0190$    & $-0.3255 \pm 0.0139$  \\
      C     & $9.4590\pm 0.5789$ & $3.9314\pm 0.0495$ &                 \                  & $8.9941\pm 0.7648$   & $4.1885 \pm 0.0564$    &                 \                    \\
      D     &                    \              &                 \                 &                 \                  &                \                    &                \                      &                 \                    \\
      E     &                     \             & $4.8071\pm 0.5403$ &                 \                  &                \                    & $6.6852 \pm 0.8549$    &                 \                     \\ \hline  
$\chi^2$&                    \multicolumn{3}{c|}{195.59+151.34=346.93}                      &                           \multicolumn{3}{c}{180.10+111.69=291.79}                \\ \hline
$\langle x q^+\rangle$ &   0.4320  &            \multicolumn{2}{c|}{0.3575}              &             0.4277             &                        \multicolumn{2}{c}{ 0.3620 }          \\ \hline\hline 
\end{tabular}
\caption{Refit to HERA~\cite{H1:2015ubc} and BCDMS~\cite{BCDMS:1989qop} data with $Q^2 \in [3.5,30000]\ {\rm GeV}^2,\ N_{\rm data} = 158 + 101 = 259$. 
The individual contributions to the total $\chi^2$ correspond to HERA and BCDMS sets, respectively.
The parameters $A$ for $x u_v$ and $xg$ are fixed by number and momentum sum rules, respectively.}
\label{tab:refit-7para}
\end{center}
\end{table*}
\begin{table*}[!htbp]
 \begin{center}
\begin{tabular}{c|ccc|ccc}
\hline\hline
              \                          &    \multicolumn{3}{c|}{$\Delta \chi^2 = 1\ (68\% \ {\rm CL})$}             &     \multicolumn{3}{c}{$\Delta \chi^2 = 2\ (95\% \ {\rm CL})$}      \\ \hline
 $(M_{AD},\epsilon)$        &      \multicolumn{3}{c|}{$(5.0,0.0205)$}                                              &     \multicolumn{3}{c}{$(5.0,0.0286)$}      \\ \hline
                 \                       &              $xg$            &          $x u_v$            &   $x \bar{u}$               &              $xg$              &          $x u_v$                &         $x \bar{u}$          \\ \hline
                A                      &              4.8556          &          4.4424              & $0.0901\pm 0.0065$ &            4.8121             &          4.4322                 & $0.0898\pm 0.0065$     \\ 
                B                      & $0.0638\pm 0.0429$ & $0.7411\pm 0.0181$ & $-0.3258\pm 0.0136$ & $0.0617\pm 0.0429$ & $0.7403\pm 0.0182$    & $-0.3261\pm 0.0137$   \\
                C                      & $8.9561\pm 0.7316$ & $4.1894\pm 0.0430$ &            \                      &  $8.9198\pm 0.7322$ & $4.1903\pm 0.0430$    &                 \                    \\
                D                      &                    \              &                 \                 &            \                      &                \                  &                \                      &                 \                    \\
                E                      &                     \             & $6.7063\pm 0.7581$  &            \                     &                \                  &  $6.7263\pm 0.7623$    &                 \                     \\ \hline  
          $\chi^2$                 &                    \multicolumn{3}{c|}{180.97 +111.82=292.79}                    &                           \multicolumn{3}{c}{181.83+111.96=293.79}                \\ \hline
$\langle x q^+\rangle$    &                0.4280         &           \multicolumn{2}{c|}{0.3617}                  &            0.4282            &                        \multicolumn{2}{c}{ 0.3614 }          \\ \hline\hline 
\end{tabular}
\caption{Fit results by including the dark photon with $\Delta \chi^2 =1$ and $\Delta \chi^2 =2$ in respect to the results with VMD ($\lambda_G = 0.897\ {\rm GeV}$) in Tab.~\ref{tab:refit-7para}.
We take the dark photon mass $m_{A_D} = 5\ {\rm GeV}$ as an example.}
\label{tab:paras7-VMD-AD}
\end{center}
\end{table*}

\begin{figure}[!h]
\includegraphics[width=\columnwidth]{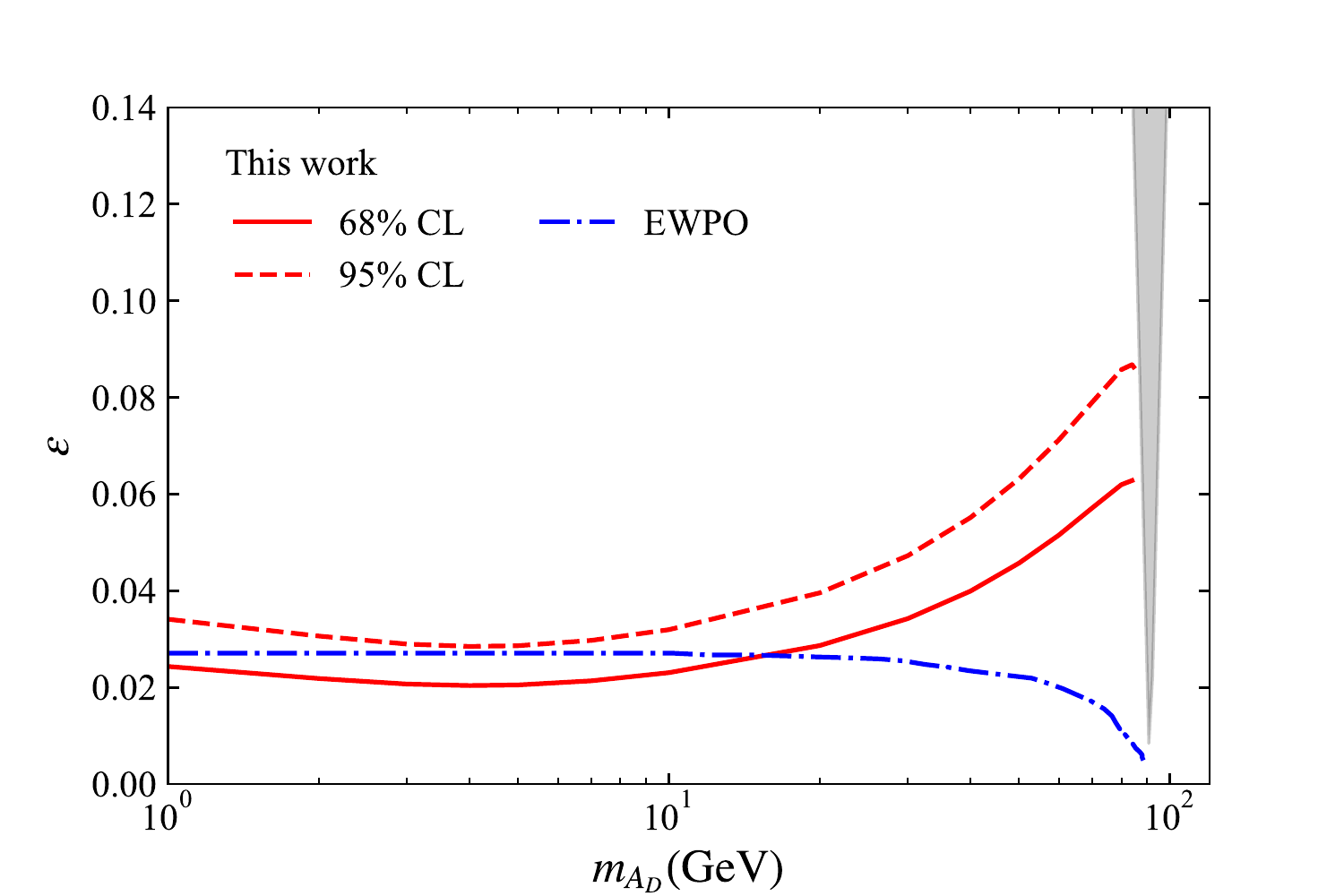}
\vspace*{-0.2cm}
\caption{The exclusion limits on the mixing parameter $\epsilon$ with $68\%$ CL (solid curve) and $95\%$ CL (dashed curve) by refitting HERA~\cite{H1:2015ubc} and BCDMS~\cite{BCDMS:1989qop} data with VMD contributions, which are compared with the EWPO constraint~\cite{Curtin:2014cca}. 
The region in grey is not accessible due to the ``eigenmass repulsion" associated with the $Z$ mass.}
\label{fig:constraints-epsilon}
\end{figure}

We then include the dark photon contribution to the proton structure function, with $\lambda_G$ fixed at $0.897\ {\rm GeV}$. 
Because of the ``eigenmass repulsion", the region of the $(\epsilon, m_{A_D})$ plane corrresponding to points near the $Z$ boson mass cannot be reached~\cite{Kribs:2020vyk}.
Therefore, we explore the parameter space of dark photon with mass up to $80\ {\rm GeV}$.
For each value of $m_{A_D}$ in the range $[1, 80]\ {\rm GeV}$, we adjust the mixing parameter $\epsilon$ and refit the DIS data. 
The minimum $\chi^2$ that the PDF fit could reach depends on $\epsilon$, $\chi^2_{\rm min} (\epsilon) - \chi^2_{\rm min} (\epsilon = 0) = \Delta \chi^2$.
The constraints on $\epsilon$ which we obtain correspond to 
$\Delta \chi^2 = 1$ ($68\%$ CL) and $\Delta \chi^2 = 2$ ($95\%$ CL),  with respect to the total $\chi^2$ of the best fit results in Tab.~\ref{tab:refit-7para}.
We show the exclusion limits on $\epsilon$ as a function of $m_{A_D}$ in Fig.~\ref{fig:constraints-epsilon}, with a more detailec summary for the particular choice 
$m_{A_D} = 5\ {\rm GeV}$ in Tab.~\ref{tab:paras7-VMD-AD}.

Our results, which correspond to constraints slightly weaker than those of Ref.~\cite{Kribs:2020vyk}, 
are competitive with the limits from electroweak precision observables~\cite{Curtin:2014cca} for the mass region $m_{A_D} \le 20 \ {\rm GeV}$.
The upper bounds on $\epsilon$ increase slightly as $m_{A_D}$ moves down towartds $1\ {\rm GeV}$, because of the interference with VMD contributions.

%

\section{Conclusion}

In this paper, we carried out an exploratory investigation of the constraints on the parameter space of the dark photon arising from studies of deep inelastic scattering (DIS).
A two-component model was applied to the proton structure function that incorporated vector meson dominance (VMD) to reproduce the correct photo-production limit, in addition to including the dark photon. The inclusion of the VMD contribution significantly improved the $\chi^2_{\rm d.o.f}$ of the fit before the addition of the dark photon. Indeed, without VMD the addition of the dark photon  actually reduced the $\chi^2_{\rm d.o.f}$. 
The exclusion limits on the kinetic mixing parameter, $\epsilon$, for the dark photon are determined by allowing variations in the goodness of fits to the extracted parton distribution functions (PDFs).
By comparison with the best fit results without the dark photon, we derived the constraints on the dark photon in the $\epsilon-m_{A_D}$ plane shown in Fig.~\ref{fig:constraints-epsilon} with both $68\%$ CL 
and $95\%$ CL.
Our results are compatible with the electroweak precision observables (EWPO) limits. 
 
The present work should be regarded as exploratory, aimed at investigating whether a full scale search based on this approach would be justified. 
We conclude that more sophisticated and accurate constraints should be obtained using a global fit analysis that goes beyond leading order.

\section*{Acknowledgements}
This work was supported by the University of Adelaide and the Australian Research Council through the Centre of Excellence for Dark Matter Particle Physics (CE200100008) and Discovery Project DP180100497 (AWT).


\end{document}